\documentclass[aps,pra,twocolumn,superscriptaddress,showpacs,floatfix]{revtex4}
\usepackage{amsmath,amssymb,amsfonts,bbm,graphicx,color}
\newcommand{\STErev}[1]{#1}
\newcommand{\ave}[1]{\langle #1 \rangle}

\def\bmsigma{{\boldsymbol{\sigma}}}
\def\bmS{{\boldsymbol{S}}}
\def\bmR{{\boldsymbol{R}}}

\def\bmOmega{{\boldsymbol{\Omega}}}
\def\e{{\rm e}}
\def\Nsq{N_{\rm sq}}
\def\Rth{R_{\rm th}}
\def\Nth{N_{\rm th}}

\def\Id{{\mathbbm I}}

\def\e{\mbox{e}}
\def\cS{{\mathcal S}}
\def\cA{{\mathcal A}}

\def\aS{a_{\rm s}}
\def\aA{a_{\rm a}}

\def\bmsigma{{\boldsymbol{\sigma}}}
\def\bmS{{\boldsymbol{S}}}
\def\bmR{{\boldsymbol{R}}}

\def\bmOmega{{\boldsymbol{\Omega}}}
\begin{document}
\title{\STErev{Full quantum state reconstruction of symmetric two-mode squeezed thermal states
via spectral homodyne detection}}
\author{Simone Cialdi}
\affiliation{Dipartimento di Fisica, Universit\`a degli Studi di
Milano, I-20133 Milano, Italy}
\affiliation{Istituto Nazionale di Fisica Nucleare, Sezione di 
Milano, Via Celoria 16, I-20133 Milan, Italy}
\author{Carmen Porto}
\affiliation{Dipartimento di Fisica, Universit\`a degli Studi di
 Milano, I-20133 Milano, Italy}
\author{Daniele Cipriani}
\affiliation{Dipartimento di Fisica, Universit\`a degli Studi di
Milano, I-20133 Milano, Italy}
\author{Stefano Olivares}
\affiliation{Dipartimento di Fisica, Universit\`a degli Studi di
Milano, I-20133 Milano, Italy}
\affiliation{Istituto Nazionale di Fisica Nucleare, Sezione di 
Milano, Via Celoria 16, I-20133 Milan, Italy}
\author{Matteo G. A. Paris}
\affiliation{Dipartimento di Fisica, Universit\`a degli Studi di
Milano, I-20133 Milano, Italy}
\affiliation{Istituto Nazionale di Fisica Nucleare, Sezione di 
Milano, Via Celoria 16, I-20133 Milan, Italy}
\date{\today}
\begin{abstract}
\STErev{
We suggest and demonstrate a scheme to reconstruct the symmetric two-mode
squeezed thermal states of spectral sideband modes from an optical parametric oscillator.
The method is based on a single homodyne detector and active stabilization of the cavity.
The measurement scheme have been successfully tested on different two-mode squeezed
thermal states, ranging from uncorrelated  coherent states to entangled states.} 
\end{abstract}
\maketitle
{\it  Introduction} -- Homodyne detection (HD) is an effective 
tool to characterize the quantum state of light in either the 
time \cite{tim1,tim2,tim3,tim4,tim5,tim6,tim7,tim8} or the frequency 
\cite{pik75,yue83,sch84,abb83,yur85,smi9xa,smi9xb,smi9xc,
vog89,dar94,mun95,sch96a,sch96b,rev03,rev09,dis1,dis2,dis3,dis4,dis5} domain.
In a spectral homodyne detector, the signal under investigation
interferes at a balanced beam splitter with a local oscillator (LO) with
frequency $\omega_0$. The two outputs undergo a photodetection process
and their photocurrents are combined leading to a photocurrent
continuously varying in time. The information about the spectral field
modes at frequencies $\omega_0 \pm \Omega$ (sidebands) is then retrieved by
electronically mixing the photocurrent with a reference signal with
frequency $\Omega$ and phase $\Psi$. Upon varying the phase $\theta$ of
the LO, we may access different field quadratures, whereas the phase
$\Psi$ can be adjusted to select the symmetric $\cS$ or antisymmetric
$\cA$ balanced combinations of the upper and lower sideband modes.  
\par
Measuring the sole modes $\cS$ and $\cA$
\STErev{through homodyne detection is not enough}
to assess the spectral correlation between the modes under investigation
\cite{bar:PRA:13} and, in turn, to fully characterize a generic quantum
state. In order to retrieve the full
information about the sidebands it has been suggested that one
should spatially separate the two modes \cite{hunt:05,hun02} or implement more
sophisticated setups \cite{bar:PRA:13,bar:PRL:13} involving resonator
detection. On the other hand, it would be desirable to have
schemes, which do not require structural modifications of the
experimental setup. In turn, this would make possible to embed more easily
diagnostic tools in interferometry \cite{som03} and 
continuous-variable-based quantum technology.
\STErev{Remarkably, in the relevant cases of interest for continuous
variable quantum information, such as squeezed state generation
by spontaneous parametric down-conversion, the correlation between the modes
vanishes, due to symmetric nature of the generated state.}
\par
\STErev{
In this Letter we suggest and demonstrate a measurement scheme
where the relevant information for the quantum state
reconstruction of symmetric spectral modes is 
obtained by using a single homodyne detector with active stabilization
through the Pound-Drever-Hall (PDH) technique \cite{PDH}.
In particular, the reconstruction is achieved by exploiting the phase coherence of
the setup, guaranteed in every step of the experiment, and two auxiliary combinations
of the sideband modes selected by setting the mixer phase at $\Psi = \pm \pi/4$.}
\begin{figure}
\includegraphics[width=0.99\columnwidth]{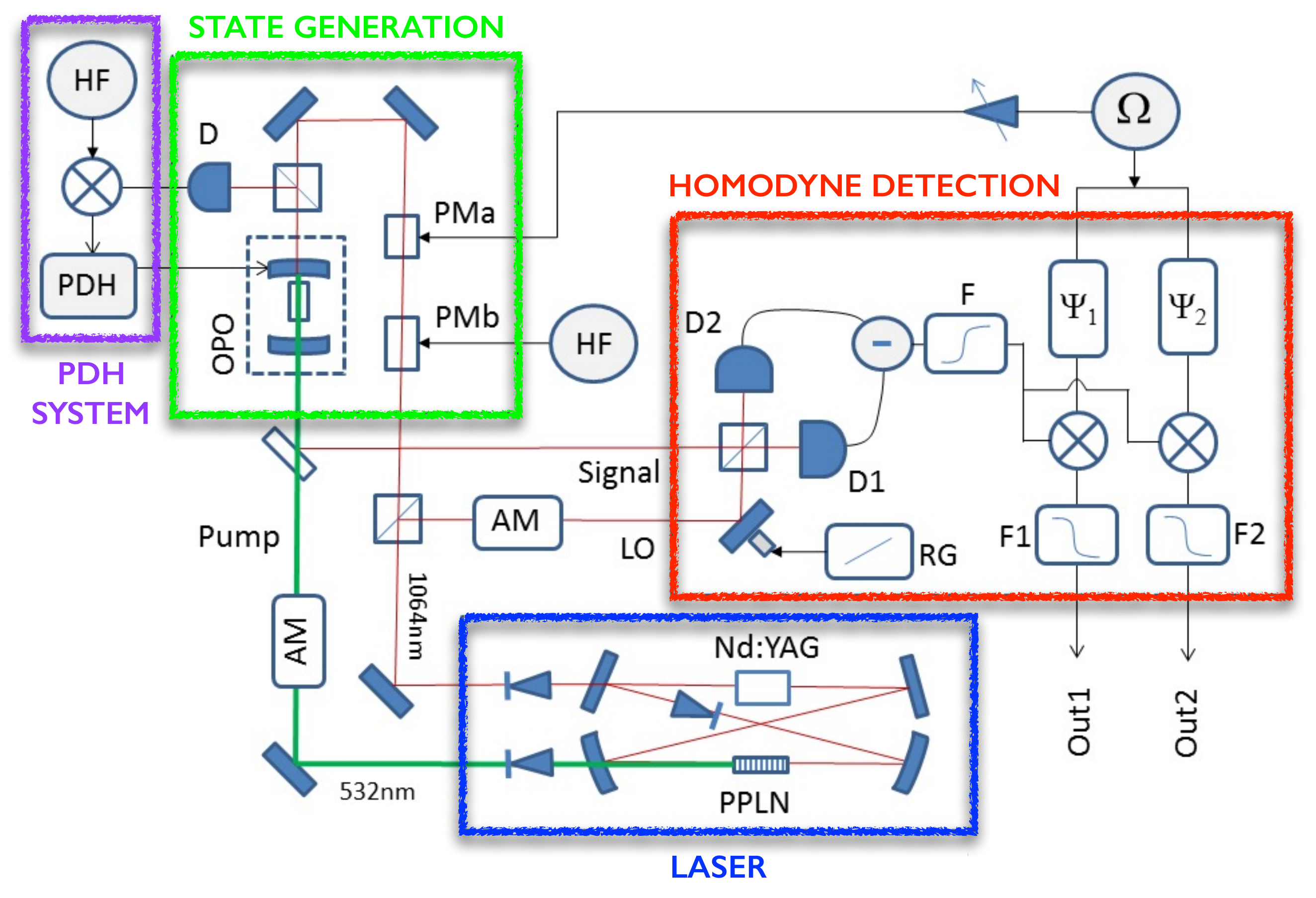}
\vspace{-0.6cm}
\caption{\label{f:scheme} Schematic diagram of the 
experimental setup. 
See the main text for details.}
\end{figure}
\par
{\it Homodyne detection and state reconstruction} -- 
A schematic diagram of our apparatus is sketched in Fig.~\ref{f:scheme}.
The principal radiation source is provided by a home made Nd:YAG Laser
($\sim$300~mW  @1064~nm and 532~nm) internally-frequency-doubled
by a periodically poled MgO:LiNbO$_3$ (PPLN in Fig.~\ref{f:scheme}). 
To obtain the single mode operation, a light diode is placed inside the laser cavity. 
One laser output (@532~nm) pumps the MgO:LiNbO$_3$ crystal of the optical
parametric oscillator (OPO) whereas the other output (@1064~nm) is sent
to a polarizing beam splitter (PBS) to generate the local oscillator
(LO) and the seed for the OPO. 
The power of the LO ($\sim$10~mW) is set
by an amplitude modulator (AM).  Two phase modulators (PMa and PMb in
Fig.~\ref{f:scheme}) generate both the sidebands used as OPO coherent
seeds and as active stabilization of the OPO cavity with the PDH technique
\cite{PDH}. For the OPO stabilization
we use a frequency of 110~MHz (HF) while the frequency $\Omega$ for the
generation of the input seed is about 3~MHz.
\STErev{This is indeed a major effort, but it will turn out to be
fundamental for the {\em full} reconstruction of the symmetric states
addressed below.}
The OPO cavity is linear with a free
spectral range (FSR) of 3300~MHz, the output mirror has a reflectivity
of 92\% and the rear mirror of 99\%. The linewidth is about 55~MHz, thus
the OPO stabilization frequency HF is well above the OPO linewidth while
the frequency $\Omega$ is well inside.  In order to actively control the
length of the OPO cavity its rear mirror is connected to a piezo that is
controlled by the signal error of the PDH apparatus.
\par
The detector consists of a 50:50 beam splitter,  two low noise detectors
and a differential amplifier based on a LMH6624 operational amplifier.
The interferometer visibility is about 95\%.  We remove  the low
frequency signal through a high-pass filter @500~kHz and then the signal
is sent to the demodulation stage. To extract the information about the
signal at frequency $\Omega$ we use an electronic setup consisting in a
phase shifter, a mixer ($\bigotimes$ in Fig.~\ref{f:scheme}) and a low-pass
filter @300~kHz. Since, as we will see in the following, we need
to measure the signal at two different orthogonal phases, $\Psi_1$ and
$\Psi_2=\Psi_1+\pi/2$, for the sake of simplicity we implemented a
double electronic setup to observe at the same time the outputs (see
Fig.~\ref{f:scheme}). Finally, the LO phase $\theta$ is scanned between 0 and
$2\pi$ by a piezo connected with a mirror before the beam splitter of
the HD. The acquisition time is 20~ms and we collect about 100~000
points by a 2~GHz oscilloscope.  
\par
If $a_0(\omega_0)$ is the photon annihilation operator of the signal
mode at the input of the HD, it is easy to show that the detected
photocurrent can be written as (note that the ``fast term'' $\omega_0$
is canceled by the presence of the LO at the same frequency) $I(t)
\propto a_0(t)\, \e^{-i\theta} + a_0^{\dag}(t)\,
\e^{i\theta}$ \cite{bachor}, where $\theta$ is the phase difference
between signal and LO and we introduced the time-dependent field
operator $a_0(t)$, that is slowly varying with respect to the
carrier at $\omega_0$, such that $a_0(t) = \e^{-i\omega_0t} \int
d\omega\, F(\omega)\,  a_0(\omega)\,\e^{-i\omega t} \equiv
\e^{-i\omega_0 t} a_0(t)$, $F(\omega)$ being the apparatus spectral
response function.
\par
To retrieve the information about the sidebands at
frequencies $\omega_0\pm \Omega$, described by the time-dependent field
operators $\hat{a}_{\pm\Omega}(t)$, we use electronic mixers set at the
frequency $\Omega$ with phase shift $\Psi$ with respect to the signal,
leading to the current $I_{\Omega}(t,\Psi) = I(t) \cos(\Omega t +
\Psi)$.  Neglecting the terms proportional to $\exp(\pm 2 i \Omega t)$
(low-pass filter), we find the following expression for operator
describing the (spectral) photocurrent $I_{\Omega}(t,\Psi) \propto
X_\theta (t,\Psi | \Omega)$, where $X_\theta(t,\Psi | \Omega) =  b
(t,\Psi | \Omega)\, \e^{-i \theta} + b^\dag (t,\Psi | \Omega)\, \e^{i
\theta}$ is the quadrature operator associated with the field operator
(note the dependence on the two sidebands):
\begin{equation}\label{transform}
b (t,\Psi | \Omega) =
\frac{a_{+\Omega}(t)\, \e^{ i \Psi} + a_{-\Omega}(t)\, \e^{-
i \Psi}}{\sqrt{2}}. 
\end{equation} Note that $\left[  b (t,\Psi |
\Omega) ,  b^\dag (t',\Psi | \Omega) \right] = \chi_{(\Delta
\omega)^{-1}}(|t-t'|)$.
\par
The interaction inside the OPO is bilinear and involves the sideband
modes $a_{\pm\Omega}$ \cite{bachor}. It is described by the
effective Hamiltonian $H_{\Omega} \propto
a_{+\Omega}^{\dag} a_{-\Omega}^{\dag} + {\rm h.c.}$, that is
a two-mode squeezing interaction. Due to the linearity of $H_{\Omega}$,
if the initial state is a coherent state or the vacuum, the generated
two-mode state $\varrho_{\Omega}$ is a Gaussian state, namely, a state
described by Gaussian Wigner functions and, thus, fully characterized by
its covariance matrix (CM) $\bmsigma_{\Omega}$ and first moment vector
$\bmR$  \cite{oli:st,weed}. It is worth noting that due to the symmetry
of $H_\Omega$, the two-sideband state is symmetric \cite{bar:PRA:13} and
can be written as $\varrho_\Omega = D_{2}(\alpha)S_{2}(\xi)
\nu_{+\Omega}(N)\otimes\nu_{-\Omega}(N)
S_{2}^{\dag}(\xi)D_{2}^{\dag}(\alpha)$, where $D_2(\alpha) =
\exp\{[\alpha(a_{+\Omega}^{\dag}+a_{-\Omega}^{\dag})
-\mbox{h.c.}]/\sqrt{2}\}$
is the symmetric displacement operator and $S_2(\alpha) = \exp(\xi
a_{+\Omega}^{\dag} a_{-\Omega}^{\dag}-\mbox{h.c.})$ the two
mode squeezing operator and $\nu_{\pm\Omega}(N)$ is the thermal state of
mode $ a_{\pm\Omega}$ with $N$ average photons \cite{oli:st}.
The state $\varrho_{\Omega}$ belongs to the so-called class of the two-mode
squeezed thermal states, generated by the application of
 $S_{2}^{\dag}(\xi)D_{2}^{\dag}(\alpha)$
to two thermal states with (in general) different energies.
In order to test our experimental setup, we acted on the OPO pump and on
the phase modulation to generate and characterize three classes of
states: the coherent ($\alpha \ne 0$ and $N,\xi=0$), the squeezed
($\xi,N \ne 0$ and $\alpha=0$) and the squeezed-coherent ($\alpha,\xi,N
\ne 0$) two-mode sideband state.  
We now consider the mode operators:
\begin{equation}\label{AS}
b (t, 0 | \Omega) \equiv \aS, \quad \mbox{and} \quad 
b  (t,\pi/2 | \Omega) \equiv \aA ,
\end{equation}
which correspond to the symmetric  ($\cS$) and antisymmetric ($\cA$)
combination of the sideband modes, respectively, and the corresponding
quadrature operators $q_{k} = X_{0}(t,\Psi_k |\Omega)$,
$p_{k} = X_{\pi/2}(t,\Psi_k |\Omega)$,  and 
$z^{\pm}_{k} = X_{\pm\pi/4}(t,\Psi_k |\Omega)$,
$k={\rm a}, {\rm s}$, with $\Psi_{\rm s} =0$ and 
$\Psi_{\rm a} = \pi/2 $. In the $\cS / \cA$ modal
basis, the first moment vector of $\varrho_{\Omega}$ 
reads $\bmR' = (\ave{q_{\rm s}},\ave{p_{\rm s}},
\ave{q_{\rm a}},\ave{p_{\rm a}})^T$ and
\STErev{
its $4 \times 4$ CM can be written in the following block-matrix form:
\begin{equation}\label{sigma:p:stationary}
\bmsigma'=
\left(\begin{array}{cc}
\bmsigma_{\rm s} & \bmsigma_\delta \\[1ex]
 \bmsigma_\delta^T &\bmsigma_{\rm a}
\end{array}
\right),\quad 
\bmsigma_\delta = \left(\begin{array}{cc}
\epsilon_q & \delta_{qp} \\[1ex]
\delta_{pq} & \epsilon_p
\end{array}
\right),
\end{equation} 
where \cite{dauria:05}:
\begin{equation}\label{sigma:p:s:m}
\bmsigma_{k} = \left(\begin{array}{cc}
\ave{q_{k}^2} - \ave{q_{k}}^2  & \frac12
\ave{(z_{k}^{+})^2-(z_{k}^{-})^2} \\[1ex]
 \frac12\ave{(z_{k}^{+})^2-(z_{k}^{-})^2}  
 & \ave{p_{k}^2} - \ave{p_{k}}^2
\end{array}
\right)
\end{equation} 
is the CM of the mode $k= {\rm a}, {\rm s}$,
$\epsilon_l= \ave{l_{\rm s}l_{\rm a}} - 
\ave{l_{\rm s}}\ave{l_{\rm a}}$,
$\delta_{l \bar{l}}= \ave{l_{\rm s} \bar{l}_{\rm a}} - 
\ave{l_{\rm s}}\ave{\bar{l}_{\rm a}}$ with $l,\bar{l}=q,p$ and $l \ne \bar{l}$.
The matrix elements of $\bmsigma_k$ can be directly measured 
from the homodyne traces of corresponding mode $a_k$ \cite{SM}, whereas 
the entries of $\bmsigma_{\delta}$ cannot.
However, the information about $\epsilon_{l}$ can be retrieved
by changing the value of the mixer phase to $\Psi=\pm\pi/4$.
In fact, it easy to show that \cite{dauria:05,dauria:PRL,buono:JOSAB}
$\epsilon_l = \frac12 \left(\ave{l_{+}^2} - \ave{l_{-}^2}\right)
 - \ave{l_{\rm s}}\ave{l_{\rm a}}$, $l=q,p$, where $q_{\pm} = X_{0}(t,\pm\pi/4 |\Omega)$
and $p_{\pm} = X_{\pi/2}(t,\pm\pi/4 |\Omega)$.}
\par
\STErev{We now focus on $\delta_{l\bar{l}}$. Given the state $\varrho_{\Omega}$, but
with different thermal contributions, these elements are equal to the
energy unbalance between the sidebands (without the contribution due to the displacement that
does not affect the CM) \cite{SM}, namely,
$\delta_{qp} = -\delta_{pq} =\Delta N_{\Omega} = 
(N_{+\Omega}-N_{-\Omega})$, which cannot
be directly accessed by the spectral homodyne detection alone.
To overcome this issue, a resonator detection method has been proposed
and demonstrated in Refs.~\cite{bar:PRA:13,bar:PRL:13}.
In our case we can exploit the error signal from the PDH stabilization
to check the symmetry of the sideband state and also to measure the presence of some
energy unbalance of the two sidebands, leading to non-vanishing $\delta_{l\bar{l}}$.
More in details, given the cavity bandwidth, the PDH error signal allows to measure the
unbalance as \cite{SM}
$\Delta N_{\Omega} = (\tau_{+\Omega} - \tau_{-\Omega})N_{\Omega}$,
where $\tau_{\pm \Omega}$
are the relative transmission coefficients associated with the two
sideband modes and $N_{\Omega} = N_{+\Omega} + N_{-\Omega}$ can be obtained
from the (reconstructed) diagonal elements of $\bmsigma_{\rm s}$ and $\bmsigma_{\rm a}$
\cite{SM,FOP}.
}
\begin{figure}[t]
\includegraphics[width=0.938\columnwidth]{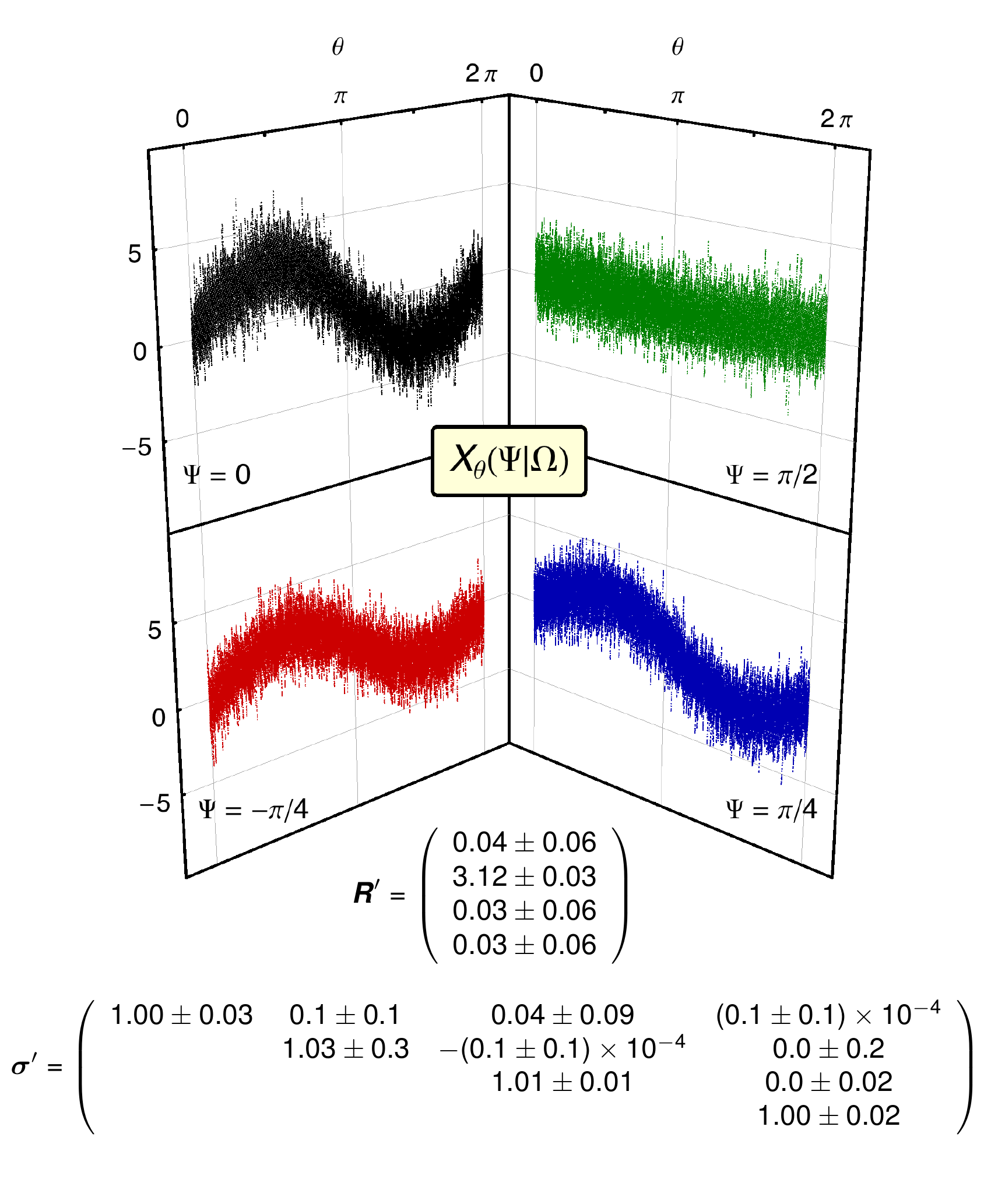}
\vspace{-0.8cm}
\caption{\label{f:trace:ch}  Homodyne traces referring to the coherent
two-mode sideband state and the reconstructed $\bmR'$ and $\bmsigma'$.
The purities of the modes $\cal S$ and $\cal A$ are $\mu_{\rm s} = 0.99
\pm 0.02$ and  $\mu_{\rm a} = 0.99 \pm 0.01$, respectively.
Only the relevant elements are shown.}
\end{figure}
\begin{figure}[t]
\includegraphics[width=0.95\columnwidth]{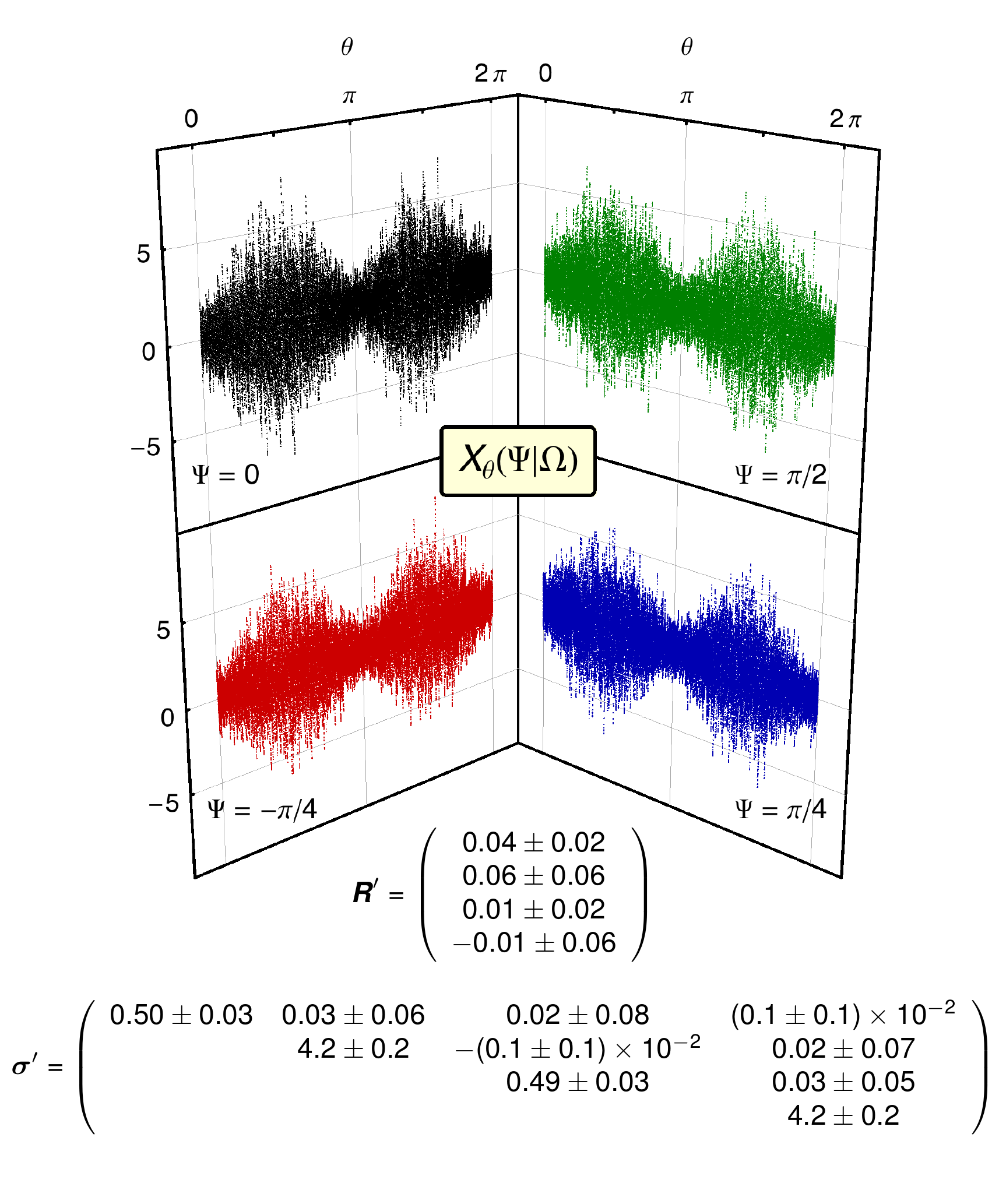}
\vspace{-0.8cm}
\caption{\label{f:trace:sq} Homodyne traces referring to the squeezed
two-mode sideband state and the reconstructed $\bmR'$ and $\bmsigma'$.
The noise reduction is $3.1 \pm 0.3$~dB for both the modes $\cal S$ and
$\cal A$, wheres their purities are $\mu_{\rm s} = 0.68 \pm 0.07$ and
$\mu_{\rm a} = 0.67 \pm 0.02$, respectively. Only the relevant elements are shown.} \end{figure}
\begin{figure}[t]
\includegraphics[width=0.95\columnwidth]{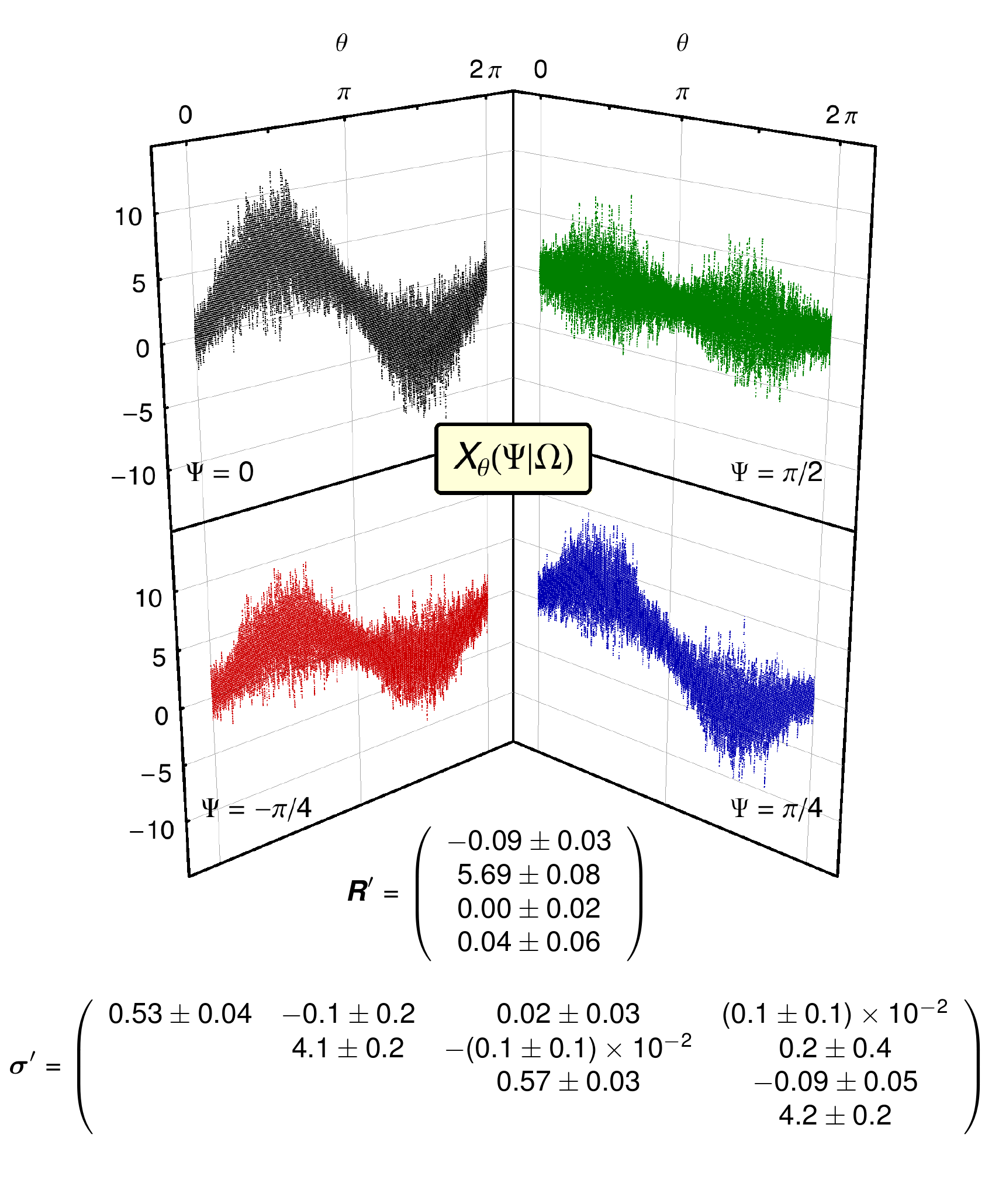}
\vspace{-0.8cm}
\caption{\label{f:trace:sq:ch} Homodyne traces referring to the
squeezed-coherent two-mode sideband state and the reconstructed $\bmR'$
and $\bmsigma'$. The noise reduction is $2.7 \pm 0.3$~dB for the $\cal
S$ mode and $2.4 \pm 0.2$~dB for the $\cal S$ mode, wheres the purities
are $\mu_{\rm s} = 0.68 \pm 0.07$ and  $\mu_{\rm a} = 0.64 \pm 0.02$,
respectively. Only the relevant elements are shown.} \end{figure}
\begin{table}[tb!]
\begin{tabular}{l}
\hline\hline
$\bullet$ {\it Two-mode coherent state:} \\[1ex]
\includegraphics[width=0.8\columnwidth]{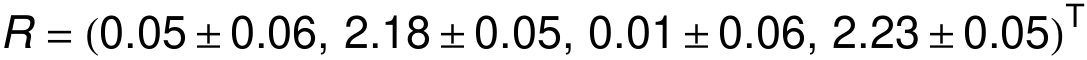} \\[1ex]
\includegraphics[width=0.8\columnwidth]{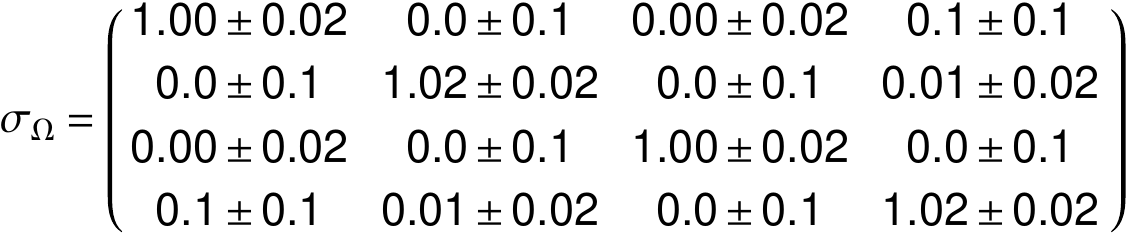} \\[1ex]
\hline\hline
$\bullet$ {\it Two-mode squeezed state:} \\[1ex]
\includegraphics[width=0.8\columnwidth]{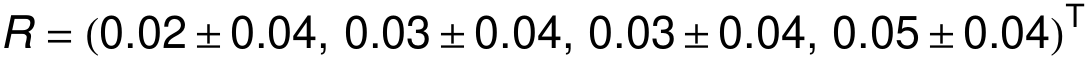} \\[1ex]
\includegraphics[width=0.8\columnwidth]{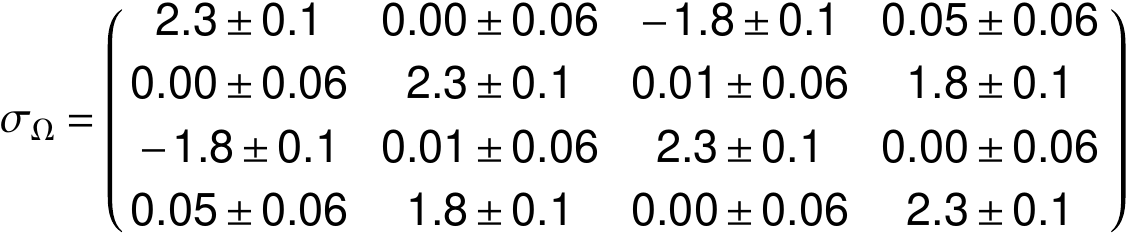}\\[1ex]
\hline\hline
$\bullet$ {\it Two-mode squeezed-coherent state:} \\[1ex]
\includegraphics[width=0.84\columnwidth]{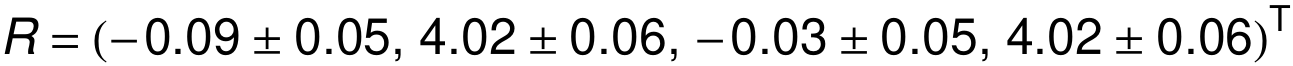} \\[1ex]
\includegraphics[width=0.8\columnwidth]{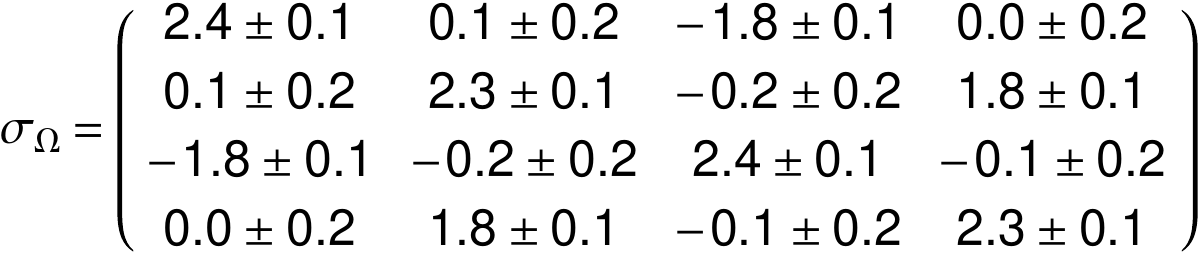}\\
\hline\hline
\end{tabular}
\caption{\label{t:sidebands} Reconstructed first moment vectors $\bmR$
and CMs $\bmsigma_\Omega$ of the two-mode sideband states
$\varrho_\Omega$ corresponding to the states of Figs.~\ref{f:trace:ch},
\ref{f:trace:sq} and \ref{f:trace:sq:ch}, respectively.} \end{table}
\par
\par
{\it Experimental results} -- Given the state $\varrho_{\Omega}$,
the full reconstruction of the CM requires the
measurement of the quadratures of modes $a_{\rm s}, a_{\rm a},$ and
$a_{\pm} = b(t,\pm\pi/4|\Omega)$.
Once the mode has been selected by choosing the suitable mixer phase
$\Psi$, the LO phase $\theta$ was scanned from $0$ to $2\pi$ to acquire
the corresponding homodyne trace.
The statistical analysis of each trace allows to reconstruct the expectation
value of the moments of the quadrature required to reconstruct the CM
$\bmsigma'$ and the first moments vector $\bmR'$.  
\par
Figures~\ref{f:trace:ch}, \ref{f:trace:sq} and \ref{f:trace:sq:ch} show
the experimental spectral homodyne traces corresponding to the coherent,
squeezed and squeezed-coherent two-mode sideband states, respectively.
In the same figures we
report the corresponding $\bmsigma'$ and $\bmR'$. All the reconstructed
$\bmsigma'$ satisfy the physical condition $\bmsigma' +i \bmOmega \geq
0$ where $\bmOmega=i \bmsigma_y \oplus \bmsigma_y$, $\bmsigma_y$ being
the Pauli matrix \cite{oli:st}. 
This implies that the modes
${\cal S}$ and ${\cal A}$ represent the same local quantum state,
namely, $\bmsigma_{\rm s}= \bmsigma_{\rm a}$: this is in agreement with
our measurement within statistical errors, as one can check from the
Figs.~\ref{f:trace:ch}, \ref{f:trace:sq} and \ref{f:trace:sq:ch}.
Furthermore, the diagonal elements of the off-diagonal blocks  are zero
within their statistical errors, in agreement with the expectation for a
factorized state of the two modes.
\par
We should now calculate the corresponding CMs in the modal basis
$\hat{a}_{+\Omega}$ and $\hat{a}_{-\Omega}$ of the upper and lower sideband,
respectively. Because of Eqs.~(\ref{AS}) we can write $\bmsigma_\Omega =
\bmS^{T}\,\bmsigma' \bmS$ and $\bmR = \bmS^{T} \bmR'$, where
\begin{equation}\label{simplettica}
\bmS=\frac{1}{\sqrt{2}} \left(\begin{array}{cc}
\Id & \Id \\
-i \bmsigma_y & i \bmsigma_y
\end{array}
\right)
\end{equation}
is the symplectic transformation associated with the mode
transformations of Eqs.~(\ref{AS}). The results are summarized in
Table~\ref{t:sidebands}. Whereas the reconstructed two-mode sideband
coherent state is indeed a product of two coherent states, the other two
reconstructed states exhibit non-classical features. In particular,
the minimum symplectic eigenvalues of the corresponding partially transposed
CMs \cite{simon,seraf} read $\tilde{\lambda} = 0.50 \pm 0.02$ and
$\tilde{\lambda} = 0.55 \pm 0.03$ for the two-mode squeezed and
squeezed-coherent state, respectively: since in both the cases
$\tilde{\lambda} < 1$, we conclude that the sideband modes are
entangled.
\par
{\it Concluding remarks} -- \STErev{In conclusion, we have presented a 
measurement scheme to fully reconstruct the class of symmetric two-mode squeezed
thermal states of spectral sideband modes, a class of states with Gaussian
Wigner functions widely exploited in continuous variable quantum technology.
The scheme is based on a homodyne detection and active stabilization,
which guarantees phase coherence in every step of the experiment, and on a
suitable analysis of the detected photocurrents.
We have shown that by properly choosing the electronic mixer phase it is
possible to select four different combinations of the upper and lower
sideband which, together with the information form the PDH error signal, 
allows to reconstruct the elements of the covariance matrix of the state under consideration.}
The scheme has been successfully demonstrated to reconstruct
both factorized and entangled sideband states.
\par
In our implementation we have used two electronic mixers and retrieved
information about two modes at a time. It is also possible to use four 
mixers and extract information about the four modes at the same time.
The method is based on a single homodyne detector
and does not involve elements outside the main detection tools of continuous 
variable optical systems. As such, our procedure is indeed a versatile
diagnostic tool, suitable to be embedded in quantum information
experiments with continuous variable systems in the spectral domain.
\par
{\it Acknowledgments} -- This work has been supported by UniMI 
through the UNIMI14 grant 15-6-3008000-609 and the H2020 Transition 
Grant 15-6-3008000-625, and by EU through the collaborative project 
QuProCS (Grant Agreement 641277). MGAP and SO thank Alberto Porzio for
useful discussions.
\vfill

\appendix
\widetext
\section*{SUPPLEMENTAL MATERIAL}
\subsection{Covariance matrix elements of the two-mode squeezed thermal state}
In this section we explicitly show how we can calculate the elements of the CM $\bmsigma'$,
given in Eq.~(3) of the main text. We recall that, according to our definitions:
\begin{equation}\label{mode:transf}
b (t, 0 | \Omega) = \frac{\hat{a}_{+\Omega}(t) + \hat{a}_{-\Omega}(t)}{\sqrt{2}} \equiv a_{\rm s},
\quad\mbox{and}\quad
b (t,\pi/2 | \Omega) = i\, \frac{\hat{a}_{+\Omega}(t) - \hat{a}_{-\Omega}(t)}{\sqrt{2}} \equiv a_{\rm a},
\end{equation}
therefore the quadrature operator
$X_\theta(t,\Psi | \Omega) = b (t, \Psi | \Omega) \, \e^{-i\theta} +
 b^{\dag} (t, \Psi | \Omega) \, \e^{i\theta}$ can be written as:
\begin{align}
X_\theta(t,\Psi | \Omega) = \cos \Psi \left[
q_s \cos \theta + p_s \sin \theta
\right] + 
\sin \Psi \left[
q_a \cos \theta + p_a \sin \theta
\right].
\end{align}
If we set $\Psi=0$, we have:
\begin{subequations}\label{s:quad}
\begin{align}
X_0(t, 0 | \Omega) & \equiv q_{\rm s} = a_{\rm s}+a_{\rm s}^\dag = 
\frac{q_{+\Omega} + q_{-\Omega}}{\sqrt{2}}
\Rightarrow
\langle q_{\rm s}^2 \rangle - \langle q_{\rm s} \rangle^2,\\
X_{\pi/2}(t, 0 | \Omega) &\equiv p_{\rm s} = i(a_{\rm s}^\dag-a_{\rm s})  = 
\frac{p_{+\Omega} + p_{-\Omega}}{\sqrt{2}}
\Rightarrow
\langle p_{\rm s}^2 \rangle - \langle p_{\rm s} \rangle^2,\\
X_{\pm\pi/4}(t, 0 | \Omega) & \equiv \frac{q_{\rm s} \pm p_{\rm s}}{\sqrt{2}}
\Rightarrow
\frac12 \langle q_{\rm s} p_{\rm s} + p_{\rm s} q_{\rm s} \rangle - 
\langle q_{\rm s} \rangle \langle p_{\rm s} \rangle,
\end{align}
\end{subequations}
for $\Psi = \pi/2$ we obtain:
\begin{subequations}\label{a:quad}
\begin{align}
X_0(t, \pi/2 | \Omega) & \equiv q_{\rm a} = a_{\rm a}+a_{\rm a}^\dag = 
\frac{p_{-\Omega} - p_{+\Omega}}{\sqrt{2}}
\Rightarrow
\langle q_{\rm a}^2 \rangle - \langle q_{\rm a} \rangle^2,\\
X_{\pi/2}(t, \pi/2 | \Omega) &\equiv p_{\rm a} = i(a_{\rm a}^\dag-a_{\rm a})  = 
\frac{q_{+\Omega} - q_{-\Omega}}{\sqrt{2}}
\Rightarrow
\langle p_{\rm a}^2 \rangle - \langle p_{\rm a} \rangle^2,\\
X_{\pm\pi/4}(t, \pi/2 | \Omega) & \equiv \frac{q_{\rm a} \pm p_{\rm a}}{\sqrt{2}}
\Rightarrow
\frac12 \langle q_{\rm a} p_{\rm a} + p_{\rm a} q_{\rm a} \rangle - 
\langle q_{\rm a} \rangle \langle p_{\rm a} \rangle,
\end{align}
\end{subequations}
On the other hand, if we set $\Psi=\pm\pi/4$ we find:
\begin{align}
X_0(t, \pm\pi/4 | \Omega) = \frac{q_{\rm a} \pm q_{\rm s}}{\sqrt{2}}, \quad
\mbox{and} \quad
X_{\pi/2}(t, \pm\pi/4 | \Omega) = \frac{p_{\rm s} \pm p_{\rm a}}{\sqrt{2}},
\end{align}
ad we have the following identities:
\begin{align}
\langle X_0^2(t, \pi/4 | \Omega) &-
X_0^2(t, -\pi/4 | \Omega) \rangle = 2 \langle q_{\rm a} q_{\rm s} \rangle \equiv
\epsilon_q, \\[1ex]
\langle X_{\pi/2}^2(t, \pi/4 | \Omega) &-
X_{\pi/2}^2(t, -\pi/4 | \Omega) \rangle = 2 \langle p_{\rm a} p_{\rm s} \rangle \equiv
\epsilon_p.
\end{align}
\par
As mentioned in the main text, it is not possible to calculate the elements $\delta_{qp}$
and $\delta_{pq}$ directly from the spectral homodyne traces \cite{bar:PRA:13}.
However, when the state under consideration is a two-mode squeezed thermal state
$\varrho_{\Omega} = D_{2}(\alpha)S_{2}(\xi)
\nu_{+\Omega}(N_1)\otimes\nu_{-\Omega}(N_2)
S_{2}^{\dag}(\xi)D_{2}^{\dag}(\alpha)$, where $D_2(\alpha) =
\exp\{[\alpha(a_{+\Omega}^{\dag}+a_{-\Omega}^{\dag})
-\mbox{h.c.}]/\sqrt{2}\}$
is the symmetric displacement operator and $S_2(\alpha) = \exp(\xi
a_{+\Omega}^{\dag} a_{-\Omega}^{\dag}-\mbox{h.c.})$ the two
mode squeezing operator and $\nu_{\pm\Omega}(N)$ is the thermal state of
mode $ a_{\pm\Omega}$ with $N$ average photons \cite{oli:st}, we can calculate
$\delta_{qp}$ and $\delta_{pq}$ as follows.
\par
Since the covariance matrix does not depend on the displacement operator, we
can assume $\alpha$. Furthermore, it is useful to introduce the following parameterization:
we define the squeezed photons per mode $\Nsq = \sinh^2 r$, the total number of thermal photons
$\Nth= N_1+N_2$, and the thermal-photon fraction $\Rth = N_1/N_{\rm th}$. Thereafter,
the energies of the two sidebands are given by:
\begin{align}
N_{+\Omega} &= \Nsq (1 + \Nth) + \Rth \Nth, \\
N_{-\Omega} &= \Nsq (1 + \Nth) + (1 - \Rth) \Nth,
\end{align}
respectively, and thus:
\begin{equation}
N_{+\Omega} + N_{-\Omega} = 2 \Nsq + \Nth (1+ 2\Nsq)\,\quad
\mbox{and} \quad
N_{+\Omega} - N_{-\Omega} = \Nth (2\Rth - 1).
\end{equation}
\par
The covariance matrix associated with $\varrho_{\Omega}$ have the following
block-matrix form:
\begin{equation}\label{CM:sb}
\bmsigma_{\Omega}=
\left(\begin{array}{cc}
A \, \Id & C \, \bmsigma_z \\[1ex]
 C \, \bmsigma_z & B \, \Id
\end{array}
\right),
\end{equation} 
where $\bmsigma_z$ is the Pauli matrix and:
\begin{subequations}
\begin{align}
A &= 1 + 2 \Nsq (1 +\Nth) + 2 \Rth \Nsq, \\
B &= 1 + 2 \Nsq (1 +\Nth) + 2 (1 - \Rth) \Nsq, \\
C &= 2 (1 +\Nth) \sqrt{\Nsq(1+\Nsq)}\,.
\end{align}
\end{subequations}
The corresponding {\em measured} covariance matrix reads (see the main text for details):
\begin{equation}\label{CM:meas}
\bmsigma'=
\left(\begin{array}{cc}
\frac12(A+B) \, \Id + C \, \bmsigma_z & (N_{+\Omega}-N_{-\Omega}) \, i \bmsigma_y \\[1ex]
(N_{+\Omega}-N_{-\Omega}) \, i \bmsigma_y & \frac12(A+B) \, \Id + C \, \bmsigma_z
\end{array}
\right),
\end{equation} 
$\bmsigma_y$ being the Pauli matrix. Note that while $\bmsigma'$ is always symmetric,
$\bmsigma_\Omega$ can be also asymmetric. 

\subsection{Retrieving the energy unbalance of the two-mode squeezed thermal state}

In order to determine the energy {\em difference} between the sidebands, we should measure the
cavity bandwidth and, by exploiting the error signal of the PDH \cite{PDH}, we can assess
the relative cavity transmission coefficients $\tau_{\pm \Omega} =
T_{\pm \Omega}/(T_{+\Omega} + T_{-\Omega})$ associated with the two sideband modes,
where $T_{\pm \Omega}$ are the actual transmission coefficients. Therefore, the energy
difference can be obtained as:
\begin{equation}
N_{+\Omega}-N_{-\Omega} =
\frac{T_{+\Omega} - T_{-\Omega}}{T_{+\Omega} + T_{-\Omega}}\,
(N_{+\Omega}+N_{-\Omega}).
\end{equation}
In general, given the covariance matrix $\bmsigma$ of a Gaussian state, the total energy can be
obtained from the sum of its diagonal elements $[\bmsigma]_{kk}$
as (without loss of generality we are still assuming the absence of the displacement):
\begin{equation}
N_{\rm tot} = \frac14 \sum_{k=1}^{4} [\bmsigma]_{kk} -1.
\end{equation}
Experimentally, we can find the total energy $N_{+\Omega}+N_{-\Omega}$ from
the first and second moments of the operators in Eqs.~(\ref{s:quad}) and in Eqs.~(\ref{a:quad}),
which are measured from the homodyne detection.

\end{document}